\documentclass[pra,twocolumn,preprintnumbers,superscriptaddress]{revtex4}
\usepackage{amsmath}    
\usepackage{graphicx}   
\usepackage{verbatim}   
\usepackage{color}      
\usepackage{subfigure}  
\usepackage{hyperref}   
\usepackage{braket}
\usepackage{amssymb}
\usepackage{xcolor}

\definecolor{purple(html/css)}{rgb}{0.5, 0.0, 0.5}
\begin{document}

\title{Quantum ratchet in disordered quantum walk}


\author{Sagnik Chakraborty}
\affiliation{The Institute of Mathematical Sciences, C. I. T. Campus, Taramani, Chennai 600113, India}
\affiliation{Homi Bhabha National Institute, Training School Complex, Anushakti Nagar, Mumbai 400094,  India}
\author{Arpan Das}
\affiliation{
Institute of Physics, P.O.: Sainik School, Bhubaneswar  751005,  India}
\affiliation{Homi Bhabha National Institute, Training School Complex, Anushakti Nagar, Mumbai 400094,  India}
\author{Arindam Mallick}
\affiliation{The Institute of Mathematical Sciences, C. I. T. Campus, Taramani, Chennai 600113, India}
\affiliation{Homi Bhabha National Institute, Training School Complex, Anushakti Nagar, Mumbai 400094,  India}
\author{C. M. Chandrashekar}
\email{chandru@imsc.res.in}
\affiliation{The Institute of Mathematical Sciences, C. I. T. Campus, Taramani, Chennai 600113, India}
\affiliation{Homi Bhabha National Institute, Training School Complex, Anushakti Nagar, Mumbai 400094,  India}

\begin{abstract}
Symmetrically evolving discrete quantum walk results in dynamic localization with zero mean displacement when the standard evolution operations are replaced by a temporal disorder evolution operation. In this work we show that the quantum ratchet action, that is, a directed transport in standard or disordered discrete-time quantum walk can be realized by introducing a pawl like effect realized by using a fixed coin operation at marked positions that is, different from the ones used for evolution at other positions. We also show that the combination of standard and disordered evolution operations can be optimized to get the mean displacement of order $\propto$ t (number of walk steps). This model of quantum ratchet in quantum walk is defined using only a set of entangling unitary operators resulting in the coherent quantum transport. 
\end{abstract}

\maketitle

\section{Introduction}


Ratchet action or ratcheting is a process of obtaining a directed transport from an unbiased source of energy produced by sources in physical systems. Ratchet, as a device allows motion in one direction while restricting the motion in the opposite direction. Ratchet action has played an important role in understanding and engineering transport phenomenon and has emerged as an active area of research interest\,\cite{00,01,02}. The literature on ratchet action dates back to the thought experiment of Feynman and Smoluchowski\,\cite{13}, where a periodic and spatially asymmetric system, in contact with a single heat bath was expected to extract work from random fluctuations (heat), apparently contradicting the second law of thermodynamics. They pointed out that, if only equilibrium fluctuations are acting, then there would be no directed motion in accordance with the second law of thermodynamics. 
Therefore, system has to be driven out of equilibrium by some additional deterministic or stochastic perturbations. These perturbations are taken to be unbiased, that is, their time, space and ensemble averages are zero. Along with that, if the spatial symmetry of the system is broken (most commonly, this is done by involving the so-called ratchet potential, which is periodic with broken spatial symmetry), then these two conditions are sufficient to have a directed transport even without a net force in a spatially periodic system. These type of ratchet systems with thermal noise are called Brownian motors. There are also other kinds of ratchets in absence of thermal noise. They are either chaotic dynamical dissipative ratchets or purely Hamiltonian ratchets. Symmetry analysis allows one to figure out necessary conditions (not sufficient) for directed transport in these systems. These symmetry analysis shows that we must have some broken symmetries to have ratchet action. Its application to understand the directed dynamics in different fields ranging from biological systems to quantum mechanical systems\,\cite{00,01,02,03,04} has continued to entice interest to explore new ratchet designs which can effectively model transport processes in various classical and quantum systems. 

In this paper, we present a model for quantum ratchet which will mimic our conventional model of ratchet as a device, in both ordered and disordered discrete-time quantum walk (DQW) system. Each step of DQW evolves a particle into superposition of discrete position space and for certain configurations of the initial state and parameter defining the evolution operators\,\cite{ref9, s1, 20},  the wave-packet spread symmetrically on both sides of the initial position. Compared to the classical random walk, the spread of the probability distribution for DQW is quadratically faster and this has resulted in the use of quantum walks as a tool for various quantum algorithms~\cite{21, 22, 23}.

However, a spreading wavepacket during the evolution of DQW has been shown to localize dynamically around the initial position in space when the parameter in the coin operation are randomly changed for each step, which we will call as temporal disorder\,\cite{14},\cite{15}. Since temporal disordered DQW (D-DQW) is homogeneous over the lattice space during each step evolution, the evolution is symmetric over time, there is no directed impulse on the wavepacket resulting in a zero mean displacement in position space. Therefore, engineering quantum ratchet in D-DQW can play an important role in engineering quantum transport in  disordered systems and to explore new models to understand directed transport in various discrete two-level quantum systems. To produce ratchet effect in D-DQW we mark a position in position space and  for that marked position we introduce a fixed coin operation different from the one used for rest of the position. The fixed coin operation at the marked position will effectively act as a pawl. This will break the symmetry in the position space and gives rise to the directed transport of the wavepacket in the desired direction. We will show that the rate of directed transport can be optimized to transport the quantum state with the mean displacement of up to the order proportional to the number of steps of the walk ($t$). Even without any optimization, the mean displacement in position space is noticeably higher over the earlier known ratchet model in DQW without disorder\,\cite{19}. One of the important aspect to note in our model is the absence of decohering noise, that is, the complete dynamics including the pawl effect is defined by the entangling unitary operators.  We show this by measuring the entanglement entropy between the position space and the particle (coin) space. Therefore, the ratchet in DQW will be very useful for coherent quantum transport. 

The paper is organized as follows. In section\,\ref{section2} we present the dynamics of DQW and show localization effect with temporal disorder. In section\,\ref{section3} we present a scheme to introduce quantum ratchet in DQW and show directed transport using different configuration of evolution. In section\,\ref{section4} we discuss the effect on entanglement between the position and particle space during the directed transport and conclude with discussion in section\,\ref{section5}.



\section{Disordered discrete-time quantum walk and dynamic localization}
\label{section2}

Discrete-time quantum walk (DQW) is a quantum analog of classical random walk in discrete space, and time \cite{aran, krp, ref9, se}. The evolution of the walk is described on a 
Hilbert space composing of a $ \text{coin Hilbert space}~  \mathcal{H}_c = \text{span}\big\{ \ket{\uparrow} = ( 1 ~~ 0)^T ,
\ket{\downarrow} = ( 0 ~~ 1)^T  \big\}  $  and a $ \text{position Hilbert space}~ \mathcal{H}_x  =  \text{span}\{\ket{x}:x \in a\mathbb{Z}\}  $.
The joint state of the system at any time $t$ is given by, 
\begin{align}
\label{state}
\ket{\Psi(t)}= \frac{1}{\sqrt{2}} \Big[ \ket{\uparrow}  \otimes  \ket{\Psi^{\uparrow}(t)}  +  \ket{\downarrow} \otimes  \ket{\Psi^{\downarrow}(t)} \Big] ~ \nonumber\\
\in  ~ \mathcal{H}_c \otimes \mathcal{H}_x 
\end{align}
where, $   \ket{\Psi^{\uparrow (\downarrow)}(t)}  = \sum\limits_x \alpha_x^{\uparrow (\downarrow)}(t) \ket{x}, ~ \alpha_x^{\uparrow (\downarrow)}(t) \in \mathbb{C} . $

Each step of walk evolution is given by a unitary evolution operator $ W( ~\vec{\theta}~  )$, which is a composition of coin operation $B(\theta)$ on a coin space followed by a coin state dependent spatial shift operation $S$ \cite{NV02}. For our choice of $ \vec{\theta} = (\theta, 0, 0) $, the coin operation is described by,
\begin{align}
 B(\theta) = e^{- i \vec{\theta}. \vec{\sigma_1}} 
 = \left( \begin{array}{cc}
          ~~ \cos\theta &  - i \sin\theta \\
             - i \sin\theta &  ~~\cos\theta\\ 
          \end{array}
 \right)
\end{align}
and the shift operation is given by, 
\begin{align} S =  \sum\limits_{x}  \ket{\uparrow}\bra{\uparrow} \otimes  \ket{x - a}\bra{x} + \ket{\downarrow}\bra{\downarrow} \otimes \ket{x + a}\bra{x}, \end{align}
where $a$ is the distance between the neighbouring position space. We can see that the operation $B(\theta)$ evolves the qubit (any two-level system) to the superposition of the basis states and the operation $S$ evolves the particle in superposition of position space. The single step DQW operator looks like,
\begin{align}
 W(\theta)\equiv S ~ \big[ ~ B(\theta)\otimes\mathbb{I}_x ~ \big],
\end{align}
where $\mathbb{I}_x = \sum\limits_x \ket{x}\bra{x}$.  Instantaneous state of the whole system, $\ket{\Psi(t+\tau)}=W(\theta)\ket{\Psi(t)}~~ \forall~ t$ or, $ \ket{\Psi(t)} = \big[ W (\theta)\big]^{ \big\lfloor \frac{t}{\tau} \big\rfloor} \ket{\Psi(0)}$ where $\tau$ is the time taken for evolving each step of the walk.

  Using the definition of translation operator, $ \sum\limits_x \ket{x \pm a}\bra{x} = e^{ \mp \frac{ i \hat{p} a}{\hbar}} = \sum\limits_k  e^{ \mp \frac{ i k a}{\hbar}} \ket{k}\bra{k},$ where $k$ is an 
  eigenstate of the momentum operator $\hat{p} : \hat{p} \ket{k} = k \ket{k}$,   we can derive the  
  effective Hamiltonian of the system according to the definition, $ W = \text{exp}\Big( - \frac{i ~ H_\text{eff} ~ \tau}{\hbar} \Big),$  
\begin{widetext}
 \begin{align}\label{ham}
  H_{\text{eff}} =  \mathbb{I}_c \otimes  \frac{\hbar \cos^{-1} \Big(  \cos \theta
  \cos \frac{ \hat{p} a}{ \hbar} \Big)  }{\tau \sqrt{ \mathbb{I}_x -  \cos^2 \theta \cos^2 \frac{\hat{p} a}{ \hbar} }}
  \bigg[ - \sigma_3 \otimes \cos \theta \sin \frac{ \hat{p} a}{ \hbar}  
  - \sigma_2 \otimes \sin \theta \sin \frac{\hat{p} a}{ \hbar}  +  \sigma_1 \otimes \sin \theta \cos \frac{\hat{p} a}{\hbar}  \bigg]
 \end{align}\end{widetext}
 where,~~$\mathbb{I}_c = \ket{\uparrow} \bra{\uparrow} +  \ket{\downarrow} \bra{\downarrow}$. Effective energy eigenvalues $E^{\pm}_{k}$ of the Hamiltonian in Eq.\,(\ref{ham}) are,
$$ E^{\pm}_{k} =  \pm \frac{\hbar}{\tau} \cos^{-1} \bigg(  \cos \theta
  \cos \frac{ k a}{ \hbar} \bigg) ~ \forall~ k \in \bigg[ - \frac{\pi \hbar}{a}, \frac{\pi \hbar}{a}~ \bigg]. $$
The group velocity is calculated as,  
\begin{equation}
\label{grpvel}
 v_g^{\pm}(\theta, k) = \frac{\partial E_k^{\pm} }{ \partial k }  =  \pm \frac{a}{\tau}  \frac{ \cos \theta
  \sin \frac{ k a}{ \hbar}}{ \sqrt{ 1 -  \big(  \cos \theta
  \cos \frac{ k a}{ \hbar}  \big)^2} }.
\end{equation} 
The group velocity can be related to the probability distribution of the walk\,\cite{KP09}. From the group velocity of single step effective Hamiltonian a maximum spread of the wavepacket in the position space can be obtained by multiplying it with the number of walk steps $t$, $|v_g^{\pm}(\theta, k)\, t|$\,\cite{NV02, 20}.

  \begin{figure}[tb]
  \includegraphics[width=0.98\linewidth]{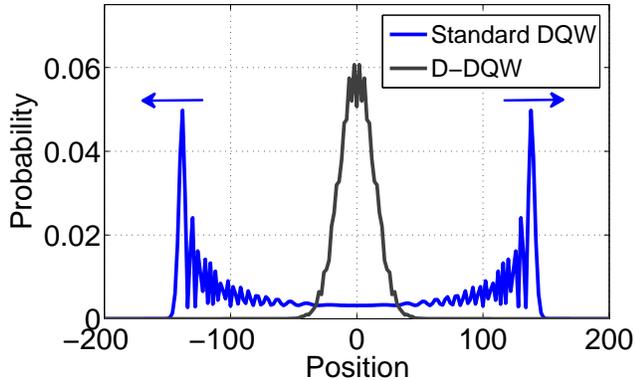}
  \caption{Probability distribution of standard DQW (with $\theta = \pi/4$) and DQW with temporal disorder after 200 steps of evolution. 
  The initial state chosen for both the evolution, $\ket{\Psi(0)} = \frac{1}{\sqrt{2}} \big(\ket{\uparrow} + \ket{\downarrow} \big) \otimes \ket{x = 0}$. 
  For standard DQW, symmetric spread of wavepacket in both the direction is seen whereas, for  D-DQW the wavepacket is localized around the origin. \label{fig:StdDisQW}}
\end{figure}

It was first predicted by Anderson\,\cite{16} that potential energy disorder can cause localization in crystals. Later it was experimentally shown to be true and theoretically verified in a variety of systems\,\cite{ref1, ref2, ref3, 17, ref4, ref5}.
Spatial disorders actually mimics spatial defects present in the media. If we consider the defects to be random, we can easily mimic the situation by introducing some random spatial disorder and this will lead to localization.
However, defects are not always present in the medium (spatial) alone, random disturbances, which means, temporal disorders can be a source of defects in the evolution. This is taken into account by having some randomly chosen temporal disturbance in the evolution of the system and this also is known to gives rise to localization (dynamic). Similar to the above results, we also expect localization of the quantum wavepacket in DQW with disordered environments in both space and time\,\cite{18}.

In this work we will only consider temporal disordered DQW (D-DQW) for our study. For introducing disorder in time we randomly pick the coin parameter $\theta_t$ for each step of the walk from an i.i.d value in range  $\{0, 2\pi\}$. 
This temporal disorder gives rise to localization of the wavepacket in position space\,\cite{14},\,\cite{15},\,\cite{18}. It has been established that the localization due to temporal disorder in D-DQW are due to quantum interference and it results in enhancement of entanglement compared to the regular DQW\,\cite{15, 15a}.  Figure\,\ref{fig:StdDisQW}  depicts the comparison of probability distribution in position space between the standard DQW and D-DQW. For the coin operation $B(\theta)$, that we have chosen, we get symmetric distribution when the initial state is $\ket{\Psi(0)} = \frac{1}{\sqrt{2}} \big(\ket{\uparrow} + \ket{\downarrow} \big) \otimes \ket{x = 0}$. 
Therefore, for all our study in this work we use $\ket{\Psi(0)}$ as the initial state so that there is no biasing of the distribution  due to the initial state. The localization under temporal disorder in the figure can also be explained by the concept of group velocity of a wavepacket as discussed earlier in this section. 

In D-DQW, the temporal disorder is homogeneous in space. So, if we start with a state at the origin, $ x = 0$,
D-DQW will spread the wavepacket as much to the left as to the right of $ x = 0$. Hence although there are excitations present in the system the wave packet remains stationary (see figure\,\ref{fig:StdDisQW}).
This can be easily seen by noting that the group velocity of the wavepacket in D-DQW for each step of the walk will be same as Eq. (\ref{grpvel}) with a time dependent coin parameter $\theta_t$ in place of $\theta$  i.e,
\begin{equation}
  v_{gd} (\theta_t, k) =  \frac{a}{\tau}  \frac{ \cos \theta_t
  \sin \frac{ k a}{ \hbar}}{ \sqrt{ 1 -  \Big(  \cos \theta_t
  \cos \frac{ k a}{ \hbar}  \Big)^2} }.
\end{equation}
Therefore, the group velocity for each step of the walk will be different and the effective group velocity will be averaged group velocity $\big< v_{gd}(t)\big>_T$ over the whole period (total time) $T$. 
So, if $ \theta$ varies over it's complete period $ [ 0, 2\pi),$ randomly or following some uniform distributed function, then  averaging over the whole period, the averaged group velocity will come out to be zero, $\big< v_{gd}(t)\big>_T  = 0 $.  For any general initial state we can show that the expectation value of position over the period D-DQW is always zero as distribution is always symmetric around the initial position and has a peak probability also around the initial position. This shows there is no transport of the wave-packet in D-DQW.  For simplicity, from hereafter we will choose $a = \tau =1$.

\section{Quantum Ratchet in D-DQW}
\label{section3}

To get a directed transport in DQW we need to have some directionality in the initial state of particle or asymmetry in the coin operation~\cite{s1}. However, these biasing will not lead anywhere towards giving a directionality for D-DQW. A ratchet in DQW should be the one which will give directionality to both, DQW and D-DQW. A ratchet like transport can arises from locally broken symmetries. Therefore, here we present a method to break the spatial symmetry of the walk operator locally which will result in introducing directionality to both DQW and D-DQW. Spatial symmetry is broken by introducing a pawl like effect in the form of fixed coin operations at marked position. 

\begin{figure}[tb]
\vskip 0.2in
  \includegraphics[width=0.86\linewidth]{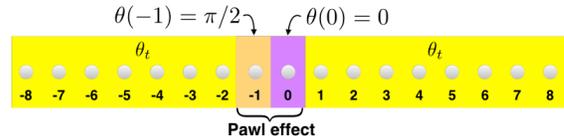}
 \caption{Schematic representation of the configuration of the evolution operation with fixed coin operations at marked positions to produce pawl like effect in DQW. \label{fig:SchematicPawl}}
\end{figure}

In figure\,\ref{fig:SchematicPawl}, we show the schematic of one possible configuration of introducing pawl like effect on the spatial position. The coin parameter, $\theta$  at $x= -1$ and $x = 0$ are chosen to be equal to $\frac{\pi}{2}$ and $0$, respectively and at all other positions, some constant value, $\theta = \theta_t$ is fixed.
\begin{figure}[tb]
  \includegraphics[width=0.98\linewidth]{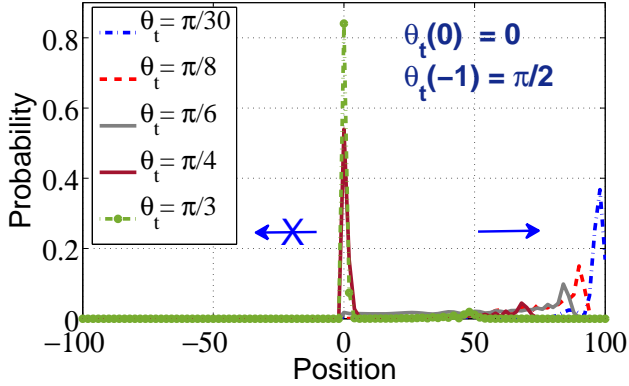}
  \caption{Probability distribution of standard DQW with $\theta$ after 100 steps of evolution. The initial state chosen for both the evolution,
  $\ket{\Psi(0)} = \frac{1}{\sqrt{2}} \big(\ket{\uparrow} + \ket{\downarrow} \big) \otimes \ket{x = 0}$. Because of the pawl like effect the wavepacket is directed towards the positive $x$ 
  direction for smaller value of $\theta$ and locked around the initial position for larger value of $\theta$. \label{fig:RaatchetRegQW}}
\end{figure}
If the initial position of the particle is at $x = 0$, after first step the state $|\uparrow\rangle \otimes |x=0\rangle$ $\rightarrow |\uparrow\rangle \otimes |x=-1\rangle$ and after second step
$|\uparrow\rangle \otimes |x=-1\rangle \rightarrow |\downarrow \rangle \otimes |x= 0\rangle$.
Similarly, if the initial state of the particle is at the position $x=-2$ the state $|\downarrow \rangle \otimes |x=-2\rangle \rightarrow |\downarrow \rangle \otimes |x=-1\rangle \rightarrow |\uparrow \rangle \otimes |x=-2\rangle$. Therefore, locally (at position $-1$), the spatial symmetry is broken.

This configuration will introduce a pawl like effect to the DQW system. Pawl like effect at marked position $x=0$ results in transport of wavepacket in increasing direction of $x$ for both standard, and disordered DQW. In figure\,\ref{fig:RaatchetRegQW}, we show the directed propagation of wavepacket for  smaller values of $\theta$ and localized component around the origin for larger value of $\theta$ when the Pawl effect is introduced. The coin operation, at $ x = -1 $ effectively blocks the spread of the wavepacket in the negative $x$ direction. Also, choosing $\theta = 0$ at $x = 0$ enhances the possibility of the rightward propagation. This can be seen from the fact that the group velocity of a wavepacket given in Eq.\,(\ref{grpvel}) is maximum for $\theta =0$ and hence for any smaller value of $\theta$
we get a delocalized wave packet in positive direction of $x$. We should note that the probability of the wavepacket at positions far from origin get significantly small with the increase in $\theta$ resulting in the wide spread of small non-zero probability at all position in the rightward direction. We will later see that this effect which is a disadvantage for wavepacket transport can be removed by introducing disorder in the system.  

\begin{figure}[tb]
  \includegraphics[width=0.98\linewidth]{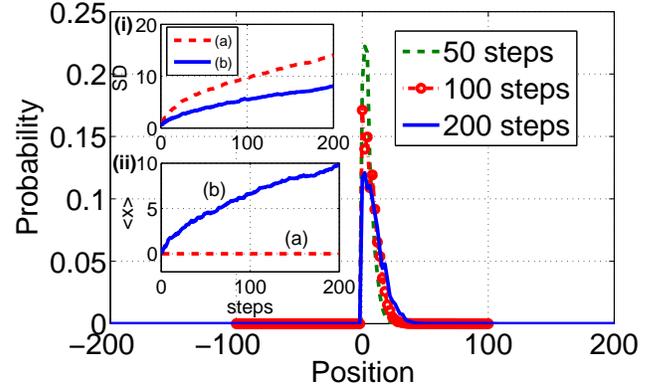}
  \caption{Probability distribution of D-DQW with pawl like effect after different number of steps. In the insets (i) and (ii) we show the standard deviation and $\langle x \rangle$ as 
  function of steps for D-DQW without pawl like effect in (a) and and with pawl like effect in (b). We can see the increase in $\langle x \rangle$ with pawl like effect along with a smaller standard deviation. $\langle x \rangle = 0$ in absence of Pawl effect.
  \label{fig:DisorderedTransport}}
\end{figure}

It is known that, random excitation or random coin operation (which does not prefer any particular direction) in D-DQW, is not able to give us directed transport in a preferred direction\,\cite{14}. But this random coin operation, helps us to localize the wavepacket, that is to say, 
standard deviation from the center of wavepacket, $ \sqrt{ \langle x^2\rangle  - \langle  x \rangle^2 }  $ becomes less and the mean displacement will be zero. Our scheme of introducing pawl effect gives directionality for a localized wavepacket. In figure\,\ref{fig:DisorderedTransport} we show the probability distribution for D-DQW after different number of steps with pawl like effect. With increase in number of steps we can note a small shift of the probability distribution to the right. To quantify the wavepacket spreading and the shift in the mean position as function of steps, we shown the standard deviation and $\langle x \rangle$ in inset (i) and (ii) of figure\,\ref{fig:DisorderedTransport} for D-DQW with (continuous line) and without pawl (dashed line) like effect. We can note that the mean value which is zero for D-DQW without pawl like effect and is shown to grow up to $10$ with pawl like effect after 200 steps. The pawl like effect also contributes to the narrowed probability distribution (smaller standard deviation). 

\begin{figure}[tb]
  \includegraphics[width=0.98\linewidth]{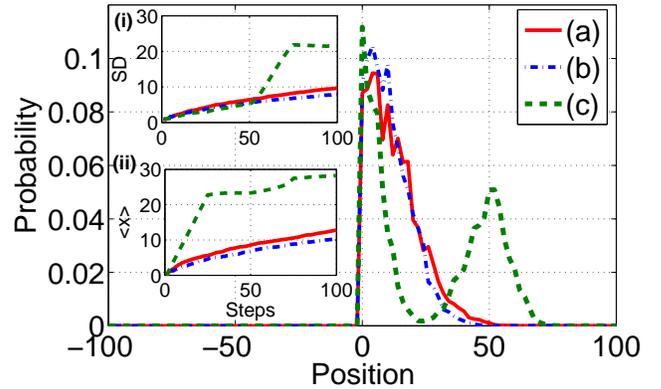}
  \caption{Probability distribution after 100 steps of DQW with pawl effect using different order of $W_1$ and $W_2$. For (a) $W_1$ with $\theta =\pi/30$ and $W_2$ are randomly picked with 1/2 probability for each step. For (b) $W_1$ with $\theta =\pi/6$ and $W_2$ are randomly picked with 1/2 probability for each step. For (c) $W_1$ with $\theta =\pi/30$ and $W_2$ are applied in order $(W_2)^{25}(W_1)^{25}(W_2)^{25}(W_1)^{25}$. In inset, SD and $\langle x \rangle$ shows that the ordered combination of operations gives a better transport.\label{fig:randomcombtheta}}
\end{figure}


From the above observation we can summarize that for standard DQW with pawl like effect, any value of $ \theta \in [0, \frac{\pi}{2}) $ gives us directional transport. Small values of $\theta$ implies higher group velocity in Eq.\,(\ref{grpvel}) and we get a localized peak with a higher $ \langle x \rangle $ value. As $\theta$ increases the peak lowers down while creating non-zero probability of occurrence at every position, which means decrease in $\langle x \rangle$. For a D-DQW with pawl like effect we will have a directed transport with a small increase in $\langle x \rangle$.

\begin{figure}[tb]
  \includegraphics[width=0.98\linewidth]{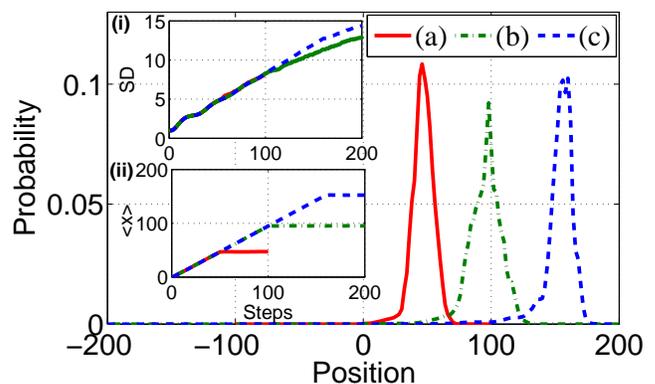}
  \caption{Probability distribution after 100 and 200 steps of DQW operation with pawl effect. For (a) $[W_2]^{50}[W_1]^{50}$ (b) $[W_2]^{100}[W_1]^{100}$ and (c) $[W_2]^{50}[W_1]^{160}$ where $\theta = \pi/30$ for $W_1$. We can see an well confined transport $\propto t/2$.
  \label{fig:conpacttransport}}
\end{figure}

Using different combination of disordered and standard DQW evolution for different interval of time we can demonstrate the control over the ratchet effect. Mathematically this means that we choose two types of walk operator, namely $W_1$ and $W_2$, and use them judiciously in a certain pattern to get a directed transport of wavepacket with significantly higher $\langle x \rangle$ value. The two walk operators we choose are $W_1 = S C_1$ and $W_2 = S C_2$ where,
\begin{widetext}
\begin{align}
 C_1 &= B \Big(\frac{\pi}{2} \Big)\otimes\ket{-1}\bra{-1} + B(0)\otimes\ket{0}\bra{0} + \sum_{x\neq -1, 0}  B(\theta)\otimes\ket{x}\bra{x}\\
 C_2 &=  B \Big(\frac{\pi}{2} \Big)\otimes\ket{-1}\bra{-1} + B(0)\otimes\ket{0}\bra{0} + \sum_{x\neq -1, 0} B \big(  \theta_t \big) \otimes\ket{x}\bra{x}.
 \end{align}
 \end{widetext}
In $C_1$, $\theta$ once fixed is retained for all $W_1$ operations whereas, in $C_2$, $\theta_t$ is chosen randomly for different time-step (different for each $W_2$). By choosing different order of applying operators $W_1$ and $W_2$ we can control and demonstrate the transport of wavepacket to give higher $\langle x \rangle$. For example, in figure\,\ref{fig:randomcombtheta} we show the probability distribution, standard deviation and $\langle x \rangle$ for an $100$ step walk with pawl like effect using different order of $W_1$ and $W_2$ operation. By inspection we can note that the use of $W_1$ has resulted in the increase of $ \langle x \rangle$ (10 after 100 steps) when compared to the evolution using only $W_2$ (6 after 100 step) as shown in figure\,\ref{fig:DisorderedTransport}. However, this transport can be further optimized, that is, for higher $\langle x \rangle$ and low standard by choosing a well ordered combination of $W_1$ and $W_2$. In figure\,\ref{fig:conpacttransport} we show the well confined probability distribution with small standard deviation and large $\langle x \rangle \propto t$ ($\langle x \rangle = t/2$).  This has been obtained by first choosing $W_1$ (with $\theta = \pi/30$) for fixed number of steps of the walk which can ensure that the wavepacket is transported at the rate proportional to number of steps. Later this transported wavepacket is confined around the position by applying $W_2$ for a finite number of steps. We should note here that for $\theta=\pi/30$ (or any other small $\theta$) the wavepacket moves away from the origin where ratchet effect introduces a broken spatial symmetry and therefore introducing randomness in walk operator later ($W_2$) will only localizes the already spread out wavepacket around its mean position. From inset (ii)  of figure\,\ref{fig:conpacttransport} we can note that the $\langle x \rangle$ continues to increase with the operation $W_1$ and starts localizing when the operation $W_2$ starts acting on the system. However, applying $W_2$ will still be effective to confine the standard deviation of the transported wavepacket. This can be seen from the inset (i) of figure\,\ref{fig:conpacttransport}  where the standard deviation is slightly larger for the wavepacket with less number of $W_2$ operation((c) in the probability distribution). 
For a the evolution with combination of operators $W_1$ followed by $W_2$,  the state of the system after time $T = T_2 + T_1$ is given by,
\begin{align}
 \ket{\Psi(T_2 + T_1)} = W_2^{T_2}~W_1^{T_1}~ \ket{\Psi(0)} 
\end{align}
and this results in $\langle x \rangle \propto \cos(\theta) T_1$ where $\theta$ is fixed coin operation in $W_1$.

Though we have presented a good combination of $W_1$ and $W_2$ which can result in an efficient transport without spreading widely over the position space, possibility of other combinations of $W$ operators can be further explored for improving the reported quantum transport with further higher $\langle x \rangle$.

\section{Entanglement entropy in ratchet}
\label{section4}

In our evolution model of ratchet like effect in quantum walks it has to be noted that the disordered introduced in the form of random coin operations is not a noise which takes the pure state to the mixed state decohering the system. Therefore, quantum interference which plays an important role in dynamics of DQW continues to play an equally important role during ratchet effect in DQW.  Here we will use the measure of entanglement to establish that the  directed transport in DQW due to ratchet effect is a coherent transport preserving the quantumness during the evolution.
Instantaneous density matrix of the system is described by, 
\begin{align}
 \rho(t) = \ket{\Psi(t)} \bra{ \Psi(t)}  
\end{align}
 \begin{figure}[tb]
  \includegraphics[width=0.98\linewidth]{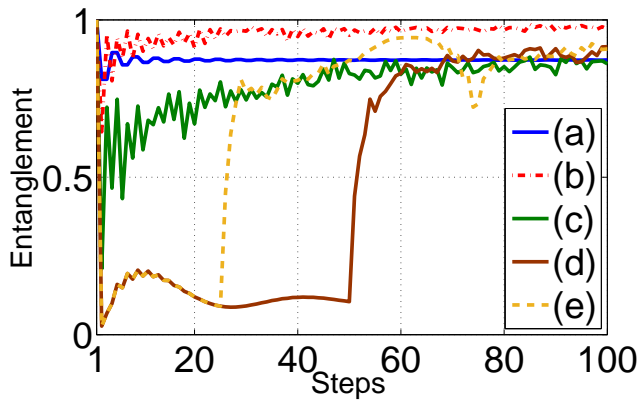}
  \caption{Entanglement with increase in number of steps for different combination of DQW with and without pawl effect. (a) and (b) are for standard standard DQW and D-DQW without pawl effect. (c) D-DQW with  like pawl effect (d) DQW with operations $W_2^{50}W_1^{50}$ where $\theta$ for $W_1$ is $\pi/30$. (e) DQW with operations $W_2^{25}W_1^{25}W_2^{25}W_1^{25}$. For all the configuration we can note that the entanglement entropy after some time reaches a values closer to 1.\label{fig:entanglement}}
\end{figure}
During DQW, the evolution of particle in superposition of position space  creates entanglement between the particle and position space. As we started with a pure initial state and whole walk operation is unitary, the system state always remain pure. Thus, entanglement entropy is enough to give the accurate measure of entanglement. 
Entanglement entropy is given by the formula, 
\begin{align}
S_e(t) =  - \text{Tr}_c \big[ \rho_c(t) ~\log \rho_c(t)  \big] 
\end{align}
where $$ \text{Tr}_x  \big[ \rho(t) \big] = \rho_c(t)$$ is a reduced density matrix obtained by partially tracing out position degree of freedom. 
In figure\,\ref{fig:entanglement} we show the entanglement entropy as a function of number of steps for different configuration of the DQW evolution with and without pawl effect.  For all the evolution with pawl effect we see that the entanglement entropy takes more time to get close to the maximum value when compared to the standard DQW and D-DQW without pawl. Therefore, after large time, the transported wavepacket will exist in a saturated entangled state with the position space ensuring the coherent transport of quantum state.

 With a demonstration of steering of random walk in ultracold atoms\,\cite{ref6}, quantum random walk of Bose-Einstein condensate (BEC) in momentum space\,\cite{ref7} and initial state dependence of quantum ratchet in BEC\,\cite{ref8}, one can anticipate realization of our ratchet model in BEC system. With the connection of DQW with Dirac cellular automaton\,\cite{qw1}, system in which dynamics can be described using Dirac equations can also be effectively explore to implement our scheme for coherent transport.  

\section{Conclusion}
\label{section5}
In this paper we have proposed a quantum ratchet model using DQW. Our configuration of fixed unitary coin operation at marked positions presents the effect of pawl. The probability distribution of the DQW, and D-DQW which spread symmetrically in position space takes a directed path when the pawl effect in the position space is introduced. We presented various configurations of the evolution operation which can be effectively used to optimize the combination of operations for maximum $\langle x \rangle \propto t$.  
We have found that the combination of evolution with sequence of fixed $\theta$ operations followed by the evolution operator with random $\theta$ can be optimized to maximize the $\langle x \rangle$ with s small standard deviation of the distribution. As the walk evolution with and without disorder, and pawl effect is completely defined using a unitary operations our transported wavepacket remain coherent during the transport and we have shown this by calculating the entanglement entropy. It would be interesting the look into the effect of decohering noise on mean displacement due to pawl effect and explore the potential application of the scheme beyond coherent quantum transport.



\end{document}